\documentclass[final]{aipproc}
\input epsf
\layoutstyle{6x9}

\begin{document}

\title{How well can we predict the total cross section at the LHC?}


\author{P V Landshoff}{
  address={Department of Applied Mathematics and Theoretical Physics \\
University of Cambridge \\
Cambridge CB3 0WA \\
England}
}

\begin{abstract}
Independently of any theory, the possibility that the large value of the Tevatron cross section claimed by CDF is correct suggests that the total cross section at the LHC may be large.
Because of the experimental and theoretical uncertainities, the best prediction is $125\pm 35$ mb.
\end{abstract}

\maketitle
This talk is based on work with Polkinghorne, Donnachie, Nachtmann and 
others going back to 1970. Further details may be found in our book\cite{book}.While theoretical understanding of 
long-range strong interactions has increased greatly since then, it
is still not good enough to allow a confident prediction of even the
value of the total cross section at the LHC. When I prepared this talk, 
I quoted $125\pm 25$ mb, but at the meeting Alan Martin predicted 90~mb.

Alan Martin's prediction is viable only if one believes that the CDF
measurement\cite{CDF94} of the $\bar pp$ cross section at the Tevatron is wrong. This
is the upper of the $\sqrt s=1800$ Gev data points shown in figure 1. 
The curves in the figure
are based on $\rho,\omega,f_2,a_2$
and soft-pomeron exchange, and they go nicely through the E710 Tevatron
data point\cite{Amo89}. At $\sqrt s=14$~TeV only the soft-pomeron term 
$21.7s^{0.0808}$ survives, giving a prediction of 101.5~mb. 

\begin{center}
\epsfxsize=.5\hsize\epsfbox[60 580 315 770]{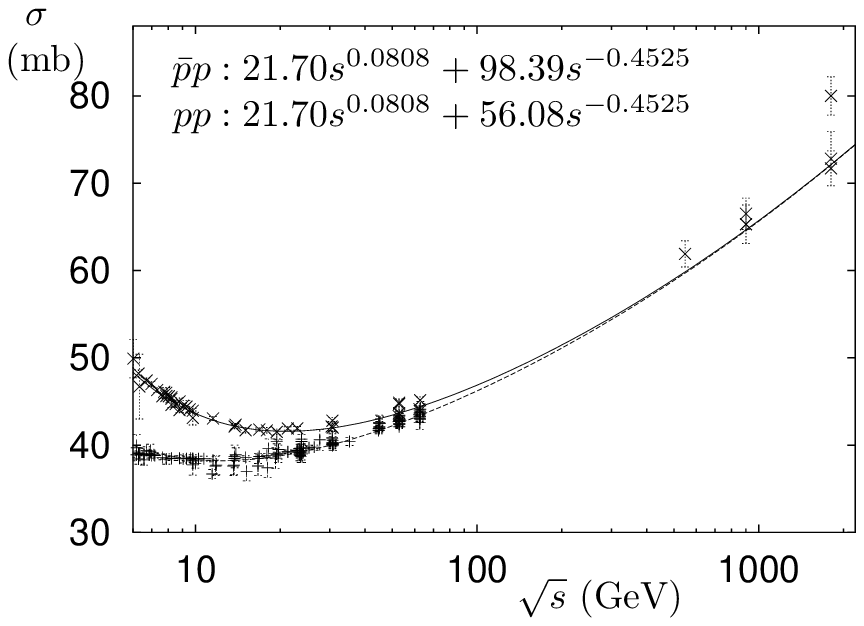}

{Figure 1: $pp$ and $\bar pp$ total cross sections}
\end{center}

A significant discovery at HERA was that soft-pomeron exchange does
not describe the rise at small $x$ of the proton structure function
$F_2(x,Q^2)$. That is, a term that behaves as $(1/x)^{\epsilon_1}$
with $\epsilon_1\approx 0.08$  is not
sufficient. As $Q^2$ increases, the data behave more and more as 
$(1/x)^{\epsilon_0}$ with $\epsilon_0\approx 0.4$. The simplest
description of the data at very small $x$ is 
\def\be{\begin{equation}}
\def\ee{\end{equation}}
\be
F_2(x,Q^2)=f_0(Q^2)x^{-\epsilon_0}+f_1(Q^2)x^{-\epsilon_1}
\label{f2}
\ee
As is well known, but is usually ignored, there are significant mathematical
difficulties in the usual perturbation-theory 
application of DGLAP evolution at small $x$.
Applying DGLAP evolution to a power fit such as (\ref{f2}) gives
differential equations for the coefficient functions $f_0(Q^2)$ and 
$f_1(Q^2)$.  However, only the one for $f_0(Q^2)$ is valid: because
$\epsilon_1$ is small, perturbation theory breaks down  for $f_1(Q^2)$.
If one extracts $f_0(Q^2)$ from fitting the small-$x$ data, it agrees
with the solution to the differential equation astonishingly well, in NLO and even
in leading order\cite{book}.

So it is natural to include also a hard-pomeron-exchange term 
$s^{\epsilon_0}$ in the fits to the $pp$ and $\bar pp$ total cross
sections\cite{factor}. Depending on how large one makes the contribution from this term,
one can make the fit go through the CDF data point, or anywhere between
the CDF and E710 points. Making it go through the CDF point leads to
a prediction of about 160~mb for the LHC total cross section.

This highlights the issue that is generally referred to as ``unitarity'', 
which can mean various things. One is the Froissart-Martin-Lukaszuk 
bound, that at large enough $s$
\font\sevenrm=cmr7
\be
\sigma^{{\sevenrm TOT}}<{\pi\over m_{\pi}^2} \log^2(s/s_0)
\ee
For reasonable values of the unknown scale $s_0$ this gives a  bound
of several barns, so it is not really relevant. A more stringent condition
is obtained by writing the elastic-scattering amplitude in so-called
eikonal form:
\be
A(s, -{\bf q}^2)
=2is\int d^2b\, e^{-i{\bf q}.{\bf b}}
\bigl(1-e^{-\chi(s,b)}\bigr)
\label{eikonal}
\ee
Then a constraint from unitarity is that Re~$\chi(s,b)\ge 0$. 

A much more difficult consequence of unitarity is that if it is
possible to exchange an object such as the soft pomeron, one must
also take account of the exchange of two or more of them. While
we know certain general features of these further exchanges, we
do not know how to make quantitative calculations. One model is
to expand the exponential in (\ref{eikonal}) as a power series:
\be
A(s,-{\bf q}^2) =2is\int d^2b\, e^{-i{\bf q}.{\bf b}} ~\Big(\chi-{{\chi^2}\over{2!}}+
{{\chi^3}\over{3!}} \dots -{{(-\chi)^n}\over{n!}}\dots\Big)
\label{cuts}
\ee
and identify the first term as the contribution from the single
pomeron exchange. The second term then has the correct features of
the exchange of two pomerons, the third term of three, and so on.
But this is only a model: there is no reason to believe that it is
correct, and various reasons to believe that it is not\cite{book}.

{\epsfxsize=0.45\textwidth\epsfbox[60 550 355 755]{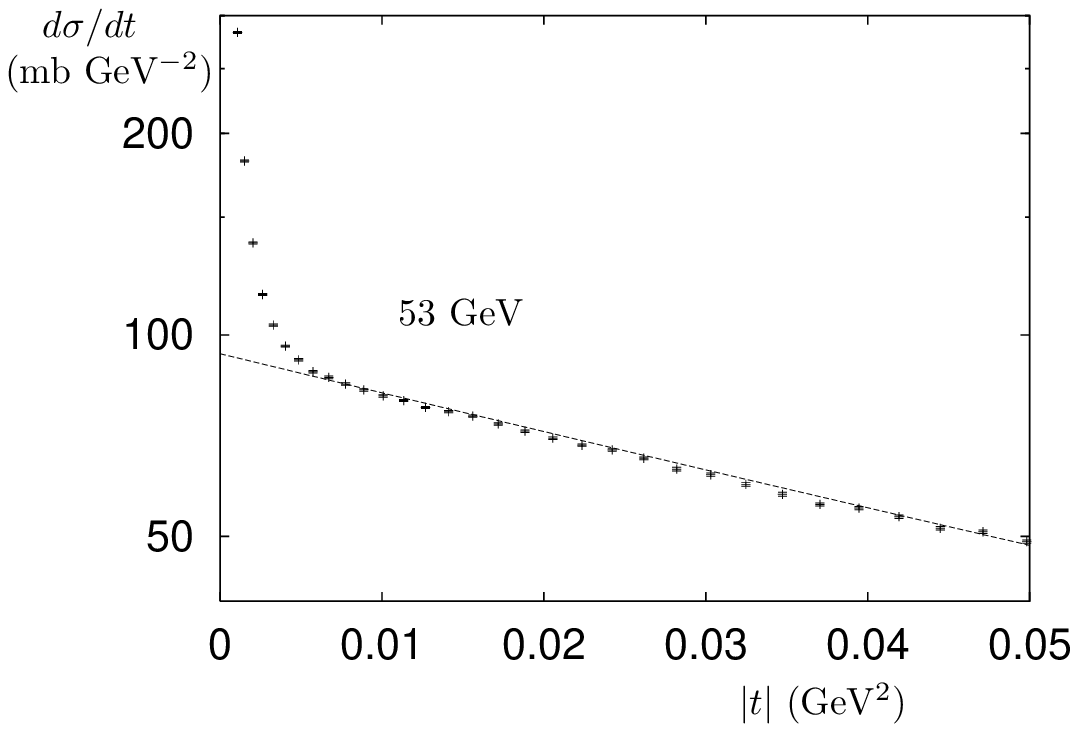}
\hfill\epsfxsize=0.45\textwidth\epsfbox[75 60 395 280]{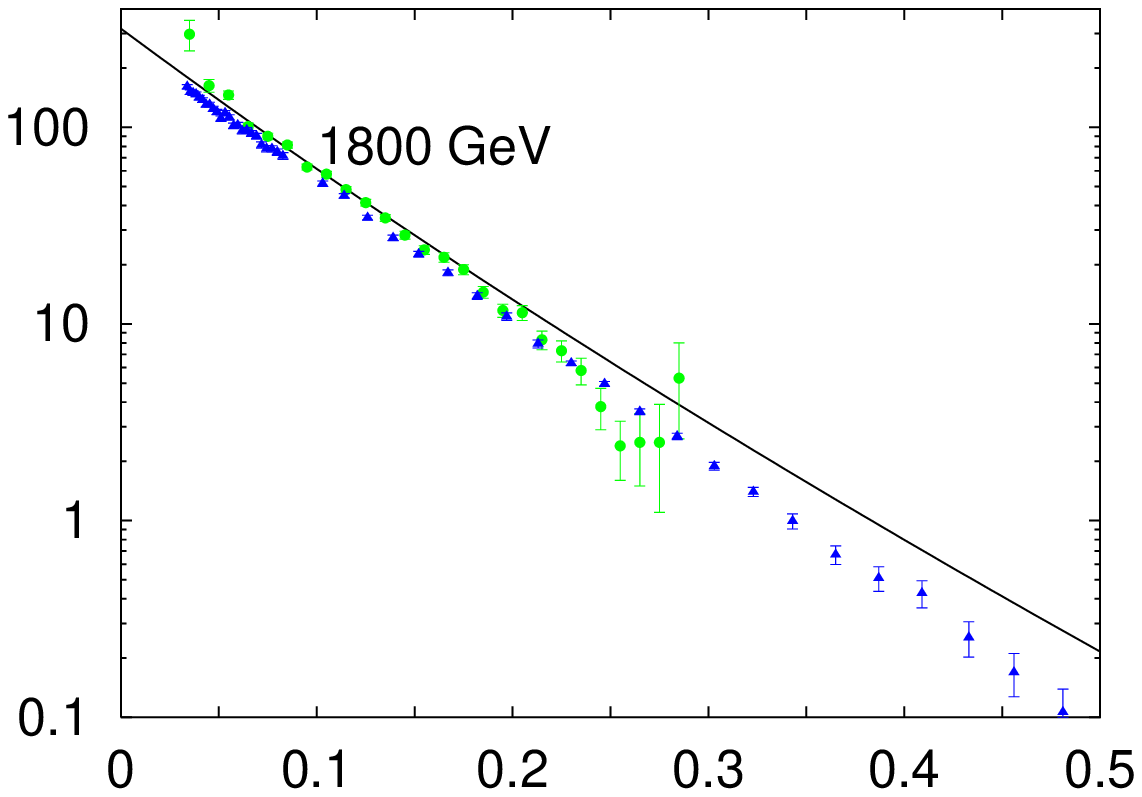}}

\begin{center} 
Figure 2: elastic scattering -- $pp$ at 53 GeV and $\bar pp$ at 1800 GeV
\end{center}

The left hand part of figure 2 shows that, beyond the Coulomb peak, single pomeron exchange gives
an excellent fit to the $pp$ elastic-scattering differential cross
section at small and medium values of $t$. 
The right-hand part of the figure shows that, at a rather higher energy,
the fit is good only at relatively small values of $t$. It is known that adding
in the contribution from the exchange of two pomerons should bend the curve 
downwards. I now describe a very crude way to calculate this.

\begin{center}
\epsfxsize=0.58\textwidth\epsfbox[88 460 475 760]{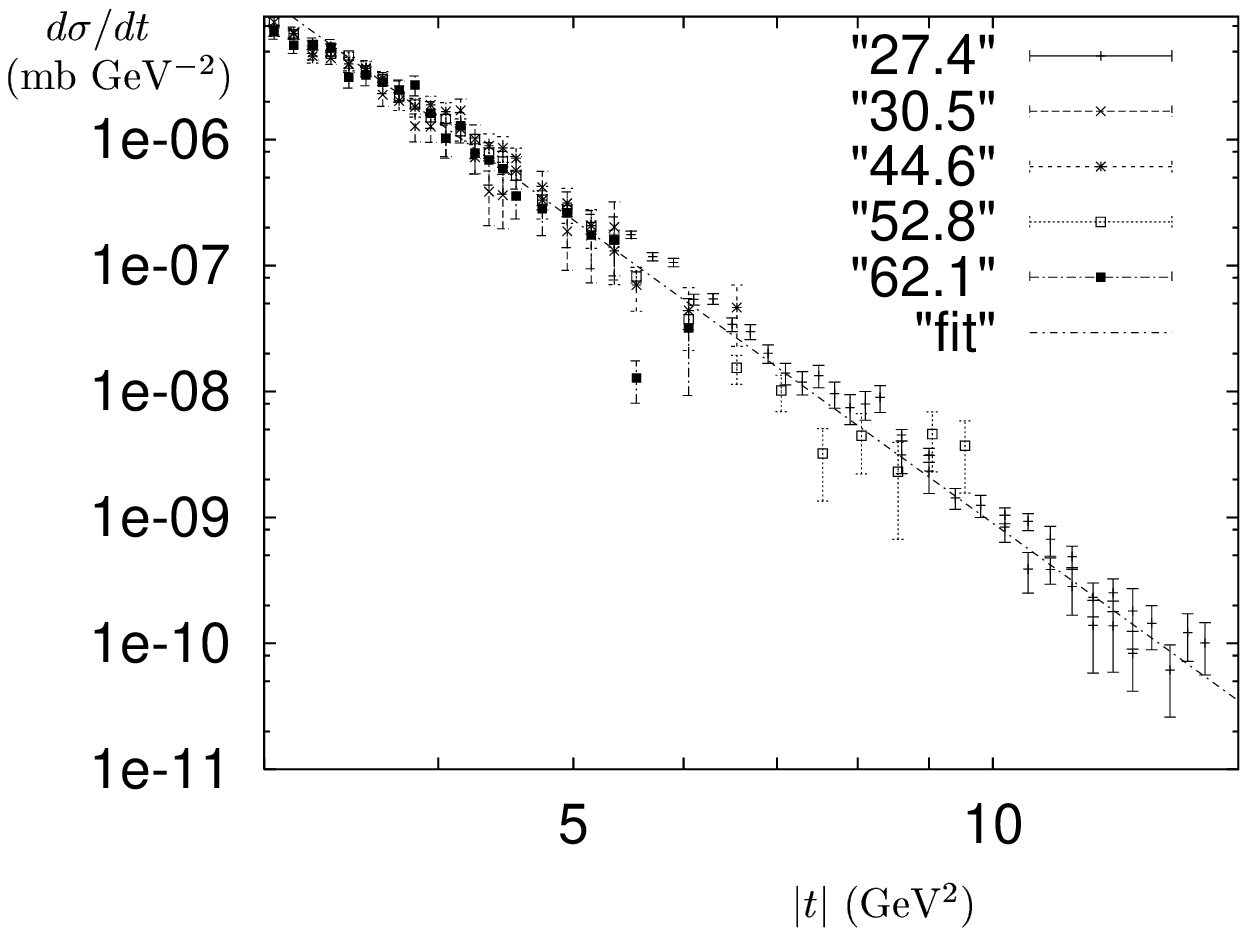}

Figure 3: $pp$ elastic scattering at large $t$. The line is $0.09~t^{-8}$.
\end{center}

First, at large $t$, the data for $pp$ elastic scattering are independent 
of energy and fit well to $d\sigma/dt\sim t^{-8}$. See figure 3. 
This behaviour is just what one gets from calculating the exchange of 3 gluons to lowest order in perturbative QCD. There is evidence that this same mechanism contributes to the creation of the dips seen in figure 4. To understand this,
note first that there are rather general principles that relate the phase of
an elastic amplitude to its energy dependence at that value of $t$. From this
one knows that, near the dip, the amplitude is neither close to being real nor
imaginary. This means that it is something of a coincidence that indeed there
is a dip: there has to be destructive interference in both the real
and the imaginary parts of the amplitude at the same value of $t$. The simplest
way to achieve this is to cancel the imaginary parts of single-pomeron and
two-pomeron exchange, and use 3-gluon exchange (which is real) to
cancel the real parts. Pomeron exchange is $C=+1$ exchange and
so does not change if we replace one of the initial protons with an
antiproton, but 3-gluon exchange changes sign because it is $C=-1$. 
So if 3-gluon exchange helps to give a dip in $pp$ scattering, it cannot do
so in $\bar pp$ scattering. And indeed experiment finds that $\bar pp$
scattering does not have a dip at $\sqrt s=53$~GeV.
\begin{center}
\centerline{\epsfxsize=0.95\hsize\epsfbox[80 560 530 770]{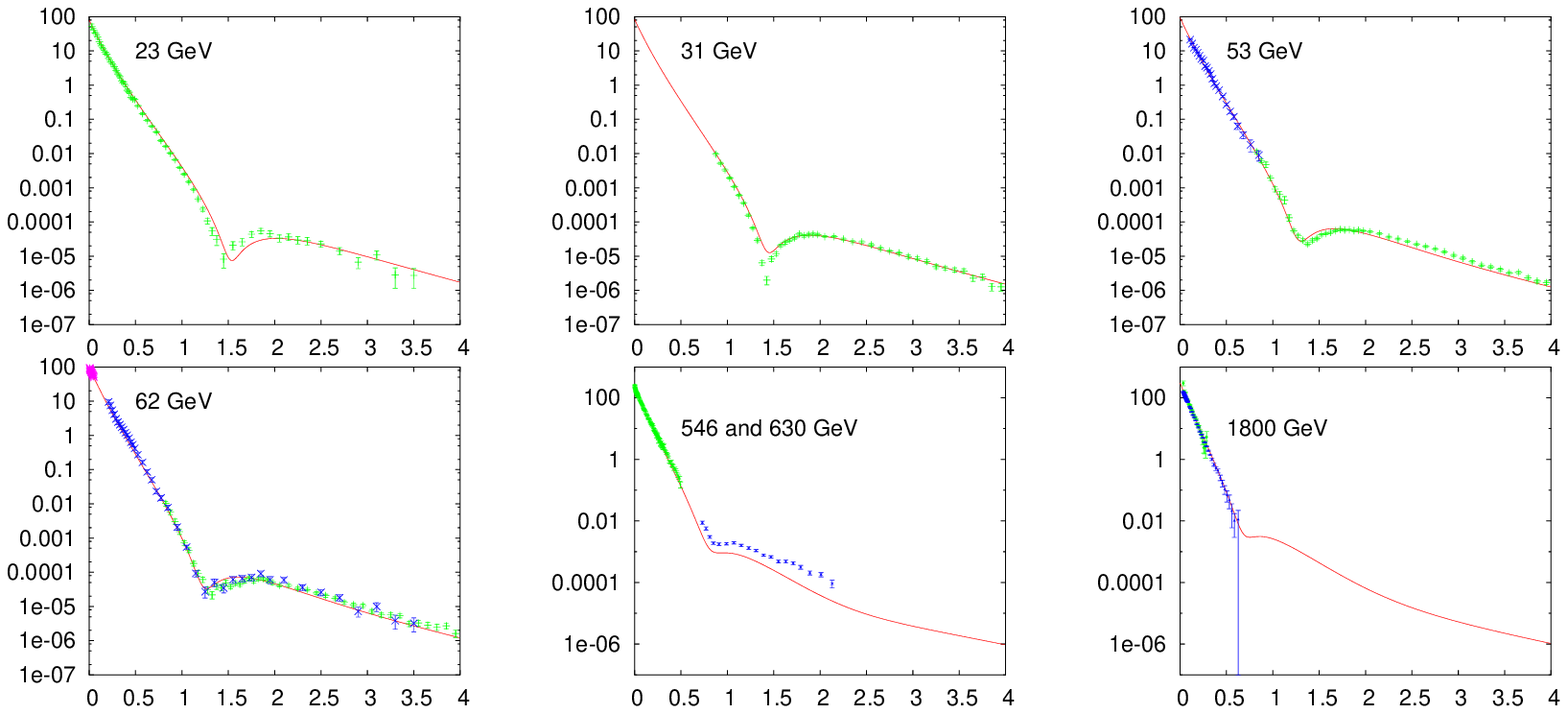}}
\vskip 3pt
Figure 4: $pp$ and $\bar pp$ elastic scattering data, with a crude model calculation.
\end{center}

So I have constructed a crude model, whose output is the curves in figure 4 and which is an adaptation of
(\ref{cuts}):
\be
A(s,-{\bf q}^2) =2is\int d^2b\, e^{-i{\bf q}.{\bf b}} ~\Big(\chi-{{\lambda\chi^2}\over{2!}} \Big)
\label{model}
\ee
I took $\chi(s,b)$ to correspond to the sum of the single exchanges of
$\rho,\omega,f_2,a_2$ and the soft and hard pomerons. The parameter
$\lambda$ determines the strength of the double exchange and is
chosen so as to cancel the imaginary part of the amplitude at the dip. 
The 3-gluon exchange term also includes a parameter that switches off its
large-$t$ behaviour, $t^{-4}$, so that it does not diverge at $t=0$. 

The result is that the power behaviour of the total cross ection from
single exchange is damped by the double exchange, and the extrapolation to
LHC energy is pulled down from 160~mb to 125~mb. Clearly, this model is very
crude, but it is the best that can be done at present.

\bibliographystyle{aipproc}   

\begin{thebibliography}{9}
\bibitem{book} A Donnachie, H G Dosch, P V Landshoff and O Nachtmann, {\sl Pomeron Physics and QCD}, Cambridge University Press (2002)
\bibitem{CDF94}
F~Abe {\em et~al},  {\bf CDF} Collaboration, Physical Review {\bf D50} (1994)
  5550
\bibitem{Amo89}
{N A Amos  {\em et~al}},  {\bf E710} Collaboration, Physical Review Letters {\bf 63} (1989) 2784
\bibitem{factor} A Donnachie and P V Landshoff, Physics Letters B595 (2004) 393

\end{thebibliography}

\end{document}